\documentclass[conference]{IEEEtran}
\IEEEoverridecommandlockouts
\usepackage{cite}
\usepackage{amsmath,amssymb,amsfonts}
\usepackage{algorithmic}
\usepackage{graphicx}
\usepackage{textcomp}
\usepackage{subfigure}
\usepackage{xcolor}
\usepackage{algorithm}
\usepackage{comment}
\def\BibTeX{{\rm B\kern-.05em{\sc i\kern-.025em b}\kern-.08em
    T\kern-.1667em\lower.7ex\hbox{E}\kern-.125emX}}
\begin{document}

\title{Compression Metadata-assisted RoI Extraction and Adaptive Inference for Efficient Video Analytics}

\author{\IEEEauthorblockN{Chengzhi Wang}
	\IEEEauthorblockA{\textit{School of Electronic Information and Communications} \\\textit{Huazhong University of Science and Technology} \\
		Wuhan, China \\
		chengzhiwang@hust.edu.cn}
	\and
	\IEEEauthorblockN{Peng Yang}
	\IEEEauthorblockA{\textit{School of Electronic Information and Communications} \\
		\textit{Huazhong University of Science and Technology} \\
		Wuhan, China\\
		yangpeng@hust.edu.cn}
	}

\maketitle

\begin{abstract}
Video analytics demand substantial computing resources, posing significant challenges in computing resource-constrained environment. In this paper, to achieve high accuracy with acceptable computational workload, we propose a cost-effective regions of interest (RoIs) extraction and adaptive inference scheme based on the informative encoding metadata. Specifically, to achieve efficient RoI-based analytics, we explore motion vectors from encoding metadata to identify RoIs in non-reference frames through morphological opening operation. Furthermore, considering the content variation of RoIs, which calls for inference by models with distinct size, we measure RoI complexity based on the bitrate allocation information from encoding metadata. Finally, we design an algorithm that prioritizes scheduling RoIs to models of the appropriate complexity, balancing accuracy and latency. Extensive experimental results show that our proposed scheme reduces latency by nearly 40\% and improves 2.2\% on average in accuracy, outperforming the latest benchmarks.
\end{abstract}


\section{Introduction}
\label{sec:intro}
With advancements in computer vision, video analytics has become essential in various applications, such as video surveillance, augmented reality, and autonomous driving \cite{zhang2024beimin, yuxin}. In particular, Deep Neural Networks (DNNs) have emerged as the preferred choice for state-of-the-art video analytics algorithms due to their remarkable accuracy. However, their high computational complexity incurs significant overhead and inference delay \cite{zhang2022maxim}. To address this, video frames are transmitted to edge servers with more computing resources for processing, referred to as edge video analytics \cite{dai2024axiomvision, chen2022context}.

Edge video analytics typically adopts two processing schemes \cite{zhang2023crossvision}. The first is frame-based analytics, where entire video frames are analyzed by pre-trained DNNs. The second is regions of interest (RoIs)-based analytics, which analyzes only RoIs extracted from frames. Peng \emph{et al.} observe that the proportion of RoIs in most videos does not exceed 25\% of the entire frame size, indicating a significant amount of redundancy in video frames \cite{peng2024arena}. Compared to frame-based analytics, RoI-based analytics can reduce the computational overhead while maintaining the accuracy.

To achieve high-accuracy RoI-based analytics with low latency, two fundamental questions are required to be considered: \textbf{how to efficiently extract RoIs} and \textbf{perform inference on these RoIs}. Essentially, the objective of RoI extraction is to accurately identify RoIs while minimizing computational overhead. RoI extraction significantly impacts analytic accuracy and overall latency, as only extracted RoIs are analyzed by DNNs. Extracting excessive RoIs increases inference overhead, while extracting insufficiently may result in missed objects. Moreover, an one-fits-all model inference scheme is not suitable for processing RoIs. The reason lies in the fact that the content of RoIs varies significantly due to camera perspective, such as variations in object sizes, etc \cite{chengzhi}. Although higher model complexity typically enhances accuracy, shallow models are sufficient for detecting the RoIs containing large objects. Thus, it is crucial to carefully design RoI inference strategies to balancing accuracy and latency.

Recent work has extensively explored various methods for extracting RoIs. Some studies employ a lightweight DNN model on the raw video at the camera \cite{cheng2023edge}. Other studies stream low-quality video to a server where more complex DNN models are deployed \cite{li2023concerto}. However, the inference accuracy of both approaches are greatly influenced by the size of DNN models and the quality of input videos. The lightweight DNN model is inherently less precise and the over-compressed video loses significant information. Thus, these factors may hinder the accurate identification of all RoIs, particularly in complex video content. Some studies tend to explore image processing techniques to extract RoIs. Wang \emph{et al.} propose a method to achieve background understanding and extraction \cite{wang2022vabus}. However, the algorithm involves many hyper-parameters specific to each video, leading to poor generalization performance in practical applications. As a result, it is necessary to design a lightweight and accurate RoI extraction method. Notably, video encoding is an indispensable part of video analytics pipeline. The video encoding encompasses a wealth of underutilized yet informative data that can identify RoIs, such as motion vectors (MVs). We argue and prove that encoding metadata can be explored to extract RoIs, thereby enhancing the efficiency of video analytics with high accuracy.

As for inference of RoIs, the diversity of objects in RoIs necessitates the deployment of different models in various scales to process RoIs effectively. Fortunately, a wide range of DNN models with varying computational complexity are available for the same detection task \cite{zhang2024beimin}. For example, the Yolov8 network for object detection offers five distinct model versions, each offering a unique trade-off between processing latency and accuracy. This allows for scheduling RoIs to an appropriate model based on specific requirements, enabling a fine-grained balance between computational overhead and accuracy. However, it is non-trivial to configure scheduling decisions. First, the scheduling model needs to be selected based on the detection difficulty of each RoI. Second, the workload of each model is required to avoid excessive latency.

In this paper, we propose an efficient edge video analytics framework for cost-effective RoI extraction and adaptive inference by models of various scales on extracted RoIs. First, MVs information of encoding metadata generated during video encoding can be used to identify RoIs within a static background. The motion regions are extracted as RoIs by parsing the encoding information and applying image processing techniques. Then, to tackle content variations among RoIs, we introduce multiple models of varying scales for fine-grained analytics. Considering the constrained processing capacity of edge nodes, our proposed scheme adaptively allocates computing resources to each model and schedules these extracted RoIs to appropriate model for processing. The main contributions of this paper can be summarized as follows.
\begin{itemize}
	\item We design an inference scheme for reference frame and non-reference frames. For the inference of non-reference frames, we propose a RoI extraction scheme based on the encoding metadata with little computational overhead.
\item We introduce multiple models of varying scales for fine grained analytics to deal with content variations among RoIs. Moreover, we propose an algorithm to allocate computing resources to these models and schedules RoIs to appropriate model scales for processing.
\item We evaluate the effectiveness of the proposed scheme in reducing computational overhead and improving accuracy through experiments on two real-world video datasets.
\end{itemize}
 The remainder of this paper is organized as follows. Section II introduces the motivation of our proposed scheme. In Section III, the detailed system model design, problem formulation and algorithm design for the problem solving are presented. Section IV shows the experimental results with other baselines. Finally, Section V summarises the paper.

\begin{figure}[t]
	\subfigure[]{
		\includegraphics[width=0.15\textwidth]{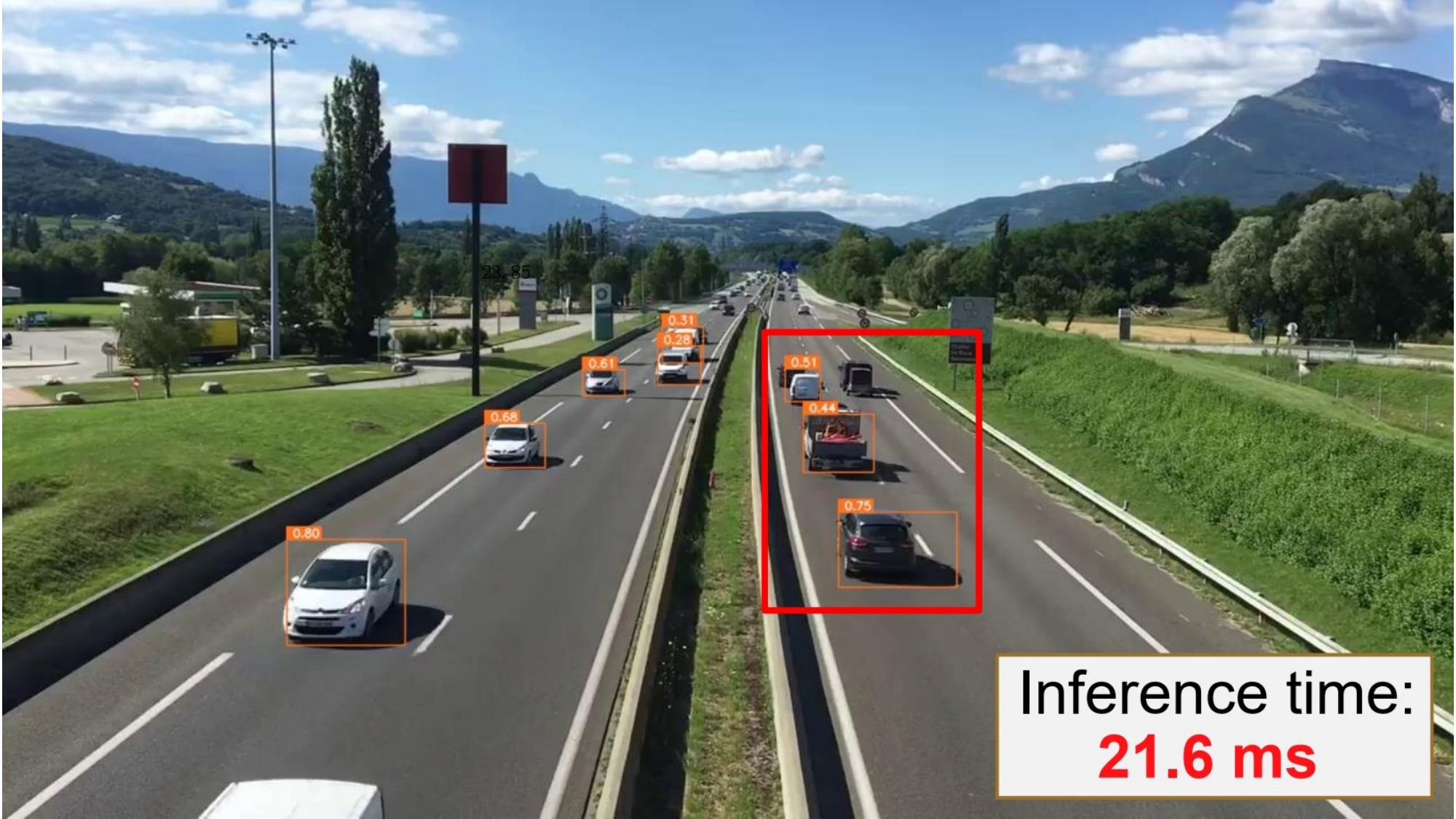}\label{fig:smallmodel}}
	\subfigure[]{
		\includegraphics[width=0.15\textwidth]{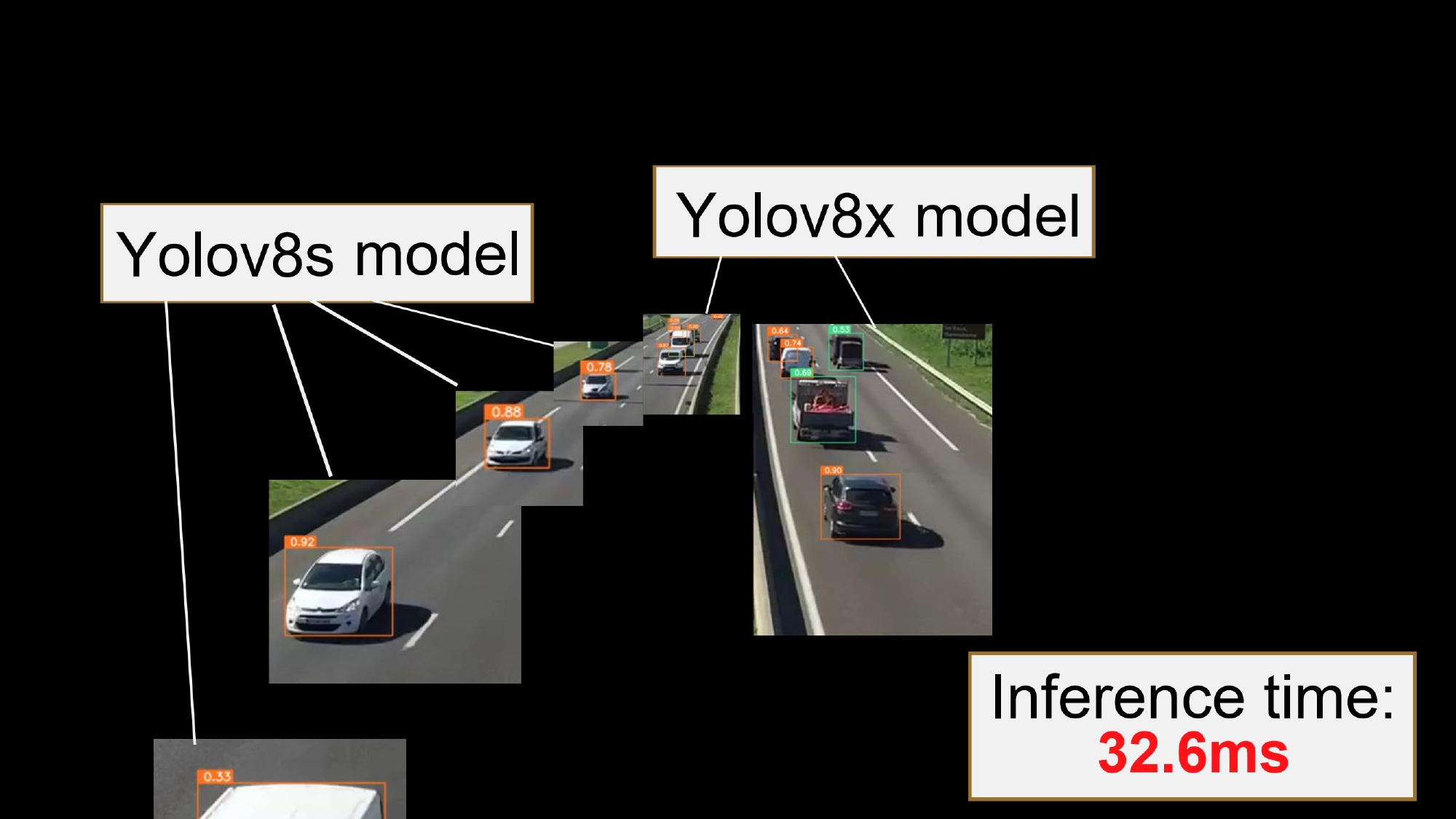}\label{fig:proposedscheme}}
	\subfigure[]{
		\includegraphics[width=0.15\textwidth]{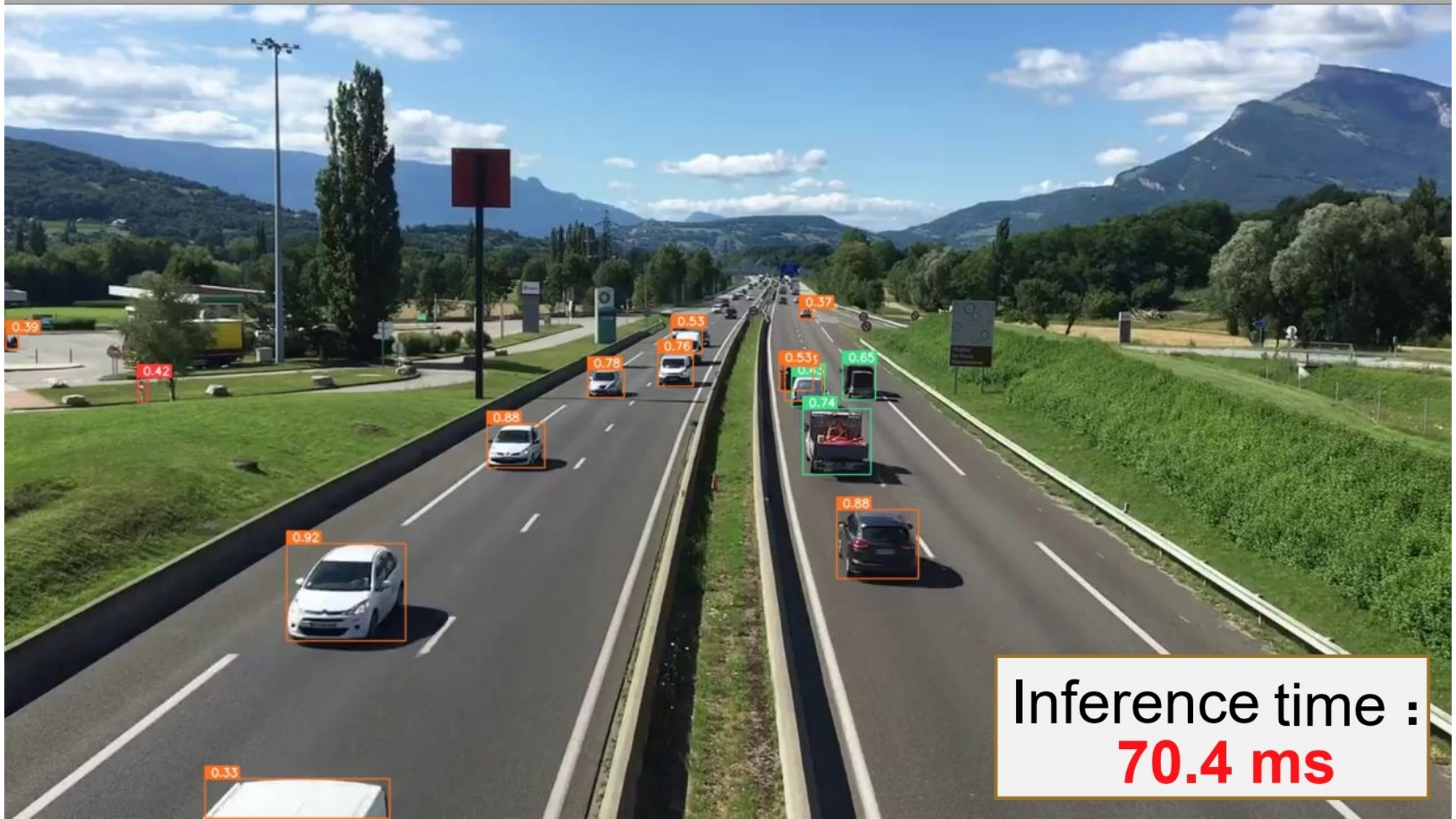}\label{fig:largemodel}}
	\vspace{-0.1in}
	\caption{A typical example of the RoI inference: inference performance of (a) the whole frame by YOLOv8s model, (b) the RoI by two models adaptively, (c) the whole frame by YOLOv8x model.}
	\vspace{-0.2in}
	\label{fig:motivation1}
\end{figure}

\begin{figure}[t]
	\centering
	\begin{minipage}[t]{0.22\textwidth}
		\centering
		\includegraphics[width=\textwidth]{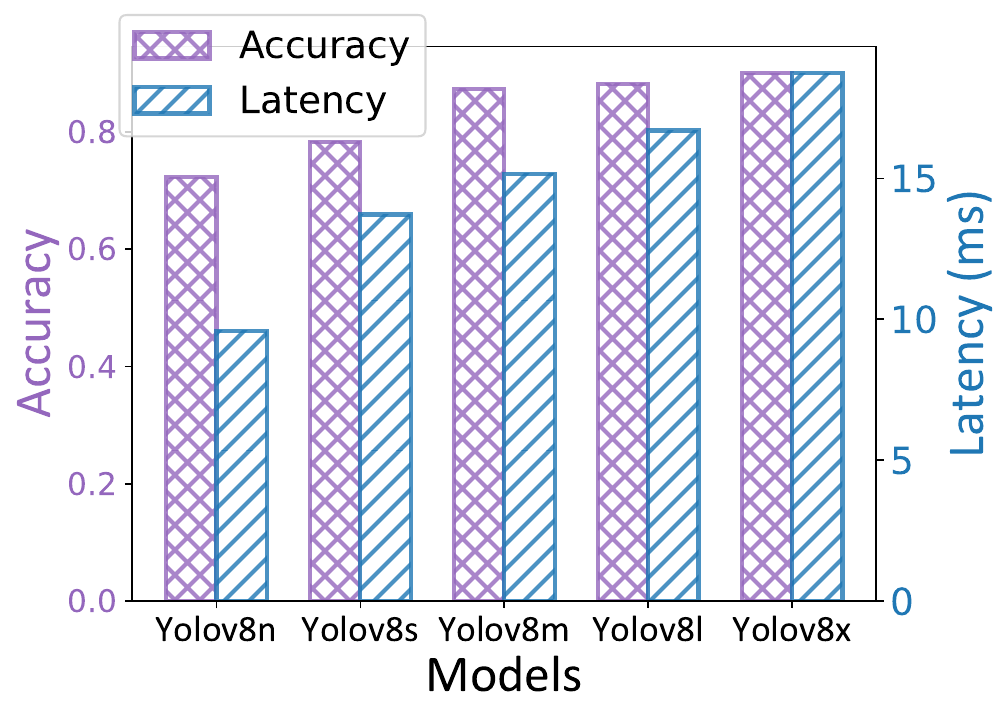}
		\vspace{-0.3in}
		\caption{Performance variation of different models in various scales with an input size of 224 * 224.}
		\label{fig:motivation2_performance}
		\vspace{-0.2in}
	\end{minipage}
	\hspace{0.1in}
	\begin{minipage}[t]{0.22\textwidth}
		\centering
		\includegraphics[width=\textwidth]{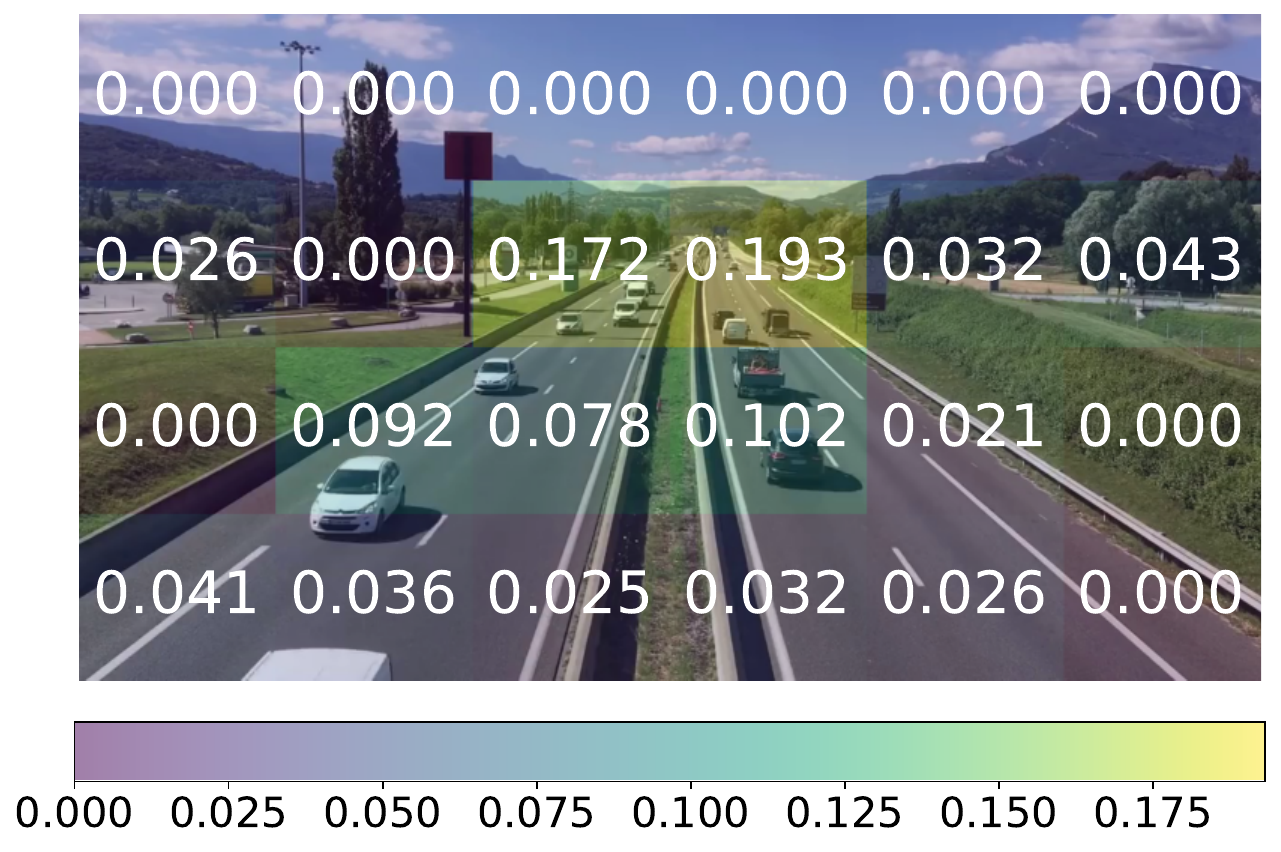}
		\vspace{-0.3in}
		\caption{Accuracy improvement differences in spatial distribution within frames.}
		\label{fig:motivation2_heatmap}
		\vspace{-0.2in}
	\end{minipage}
	\label{fig:motivation2}
\end{figure}

\section{Observations and Motivation}

 In this section, we will conduct experiments to demonstrate the significance of RoI extraction for video analytics tasks, highlighting how MVs can be effectively utilized to extract RoIs. Furthermore, we will illustrate how the spatial distribution differences in video frames motivate the use of models at varying scales for fine-grained analytics.

\subsection{RoI extraction for Video Analytics Acceleration}
 Fig. \ref{fig:motivation1} illustrates the accuracy and latency performance of a typical 720p video frame using models of different scales. The Yolov8 series, ranging from the lightweight Yolov8n to the more powerful Yolov8x, exhibit distinct accuracy-latency trade-offs, as shown in Fig. \ref{fig:motivation2_performance}. We select Yolov8s as the smaller model and Yolov8x as the larger one. The smaller model reduces overall latency but sacrifices detection accuracy, as highlighted in the red rectangle in Fig. \ref{fig:smallmodel}. In contrast, the larger model enhances accuracy at the cost of increased per-frame latency, as shown in Fig. \ref{fig:largemodel}. By adaptively inferring the extracted RoIs, we achieve a balance between accuracy and latency, as demonstrated in Fig. \ref{fig:proposedscheme}.

To efficiently extract RoIs, we observe that MVs provide critical information about inter-frame motion. Previous studies have used MVs to reuse keyframe inference results for non-inference frames \cite{yuan2023accdecoder}. However, the accuracy of reuse strategy decreases significantly as frame intervals increase, due to the cumulative effect of errors. To address this, we propose leveraging MVs from encoding information to assist in RoI extraction within non-reference frames, rather than reusing detection results directly. The approach is more robust against the impact of noise in MVs, thus maintaining better accuracy.

\subsection{Intra-frame Spatial Differences of Accuracy Improvement}
 To explore the spatial characteristics of video frames, we divide each frame into 24 blocks. These segmented blocks are detected with Yolov8s and Yolov8l models, respectively. The difference in detection accuracy between the larger model (Yolov8l) and the smaller model (Yolov8s) is termed accuracy improvement. Fig. \ref{fig:motivation2_heatmap} illustrates the average accuracy improvement across each video block, ranging from 0 to 0.2. Some blocks exhibit significant improvement, while others show none. This variation is caused by the content differences within the video frames. The blocks containing background information show no accuracy improvement, while blocks with objects show varying degrees of enhancement. This highlights the importance for careful model scale configuration for different RoIs, as increasing model complexity may introduce unnecessary computational overhead without improving accuracy for certain RoIs.

Considering the RoI extraction for video analytics acceleration and spatial distribution of video frames, we will elaborate the detailed design in the following sections.
\begin{figure*}[t]
	\centering
	\includegraphics[width=0.85\linewidth]{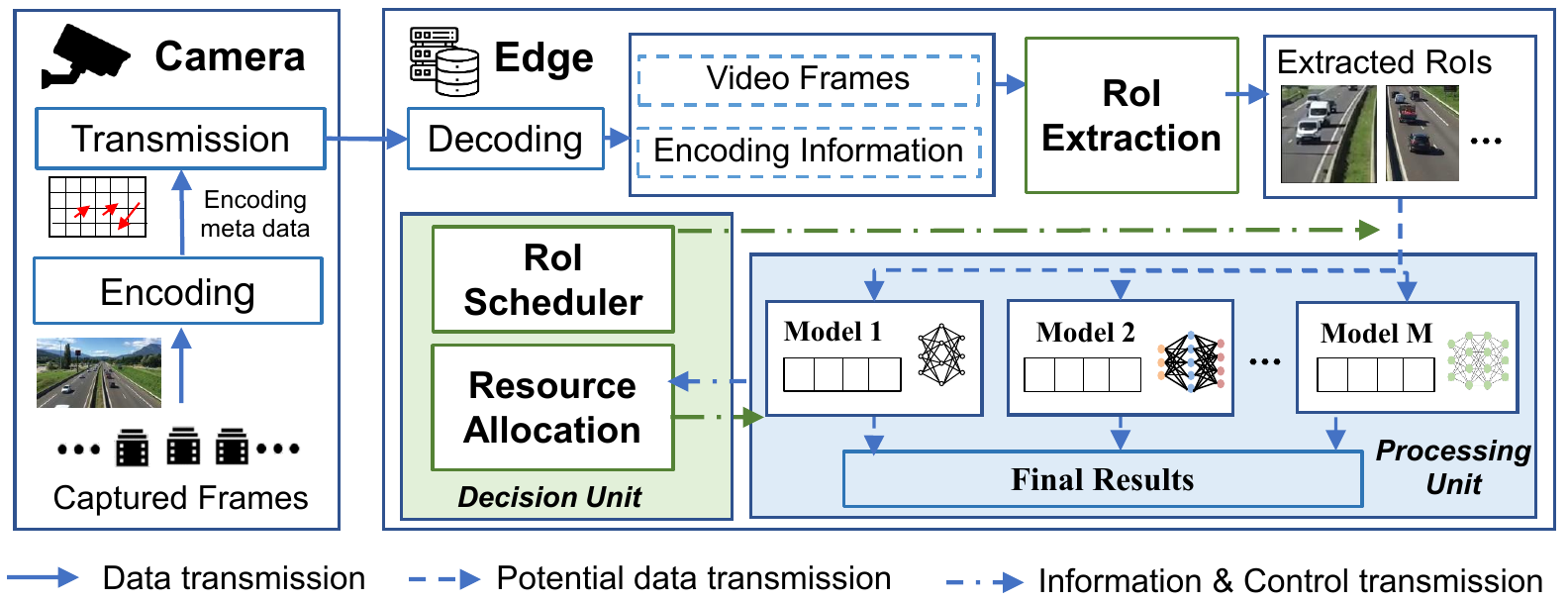}
		\vspace{-0.1in}
	\caption{\small Illustration of system model.}
	\label{fig:system_model}
	\vspace{-0.2in}
\end{figure*}
\begin{figure}[t]
	\centering
	\includegraphics[width=\linewidth]{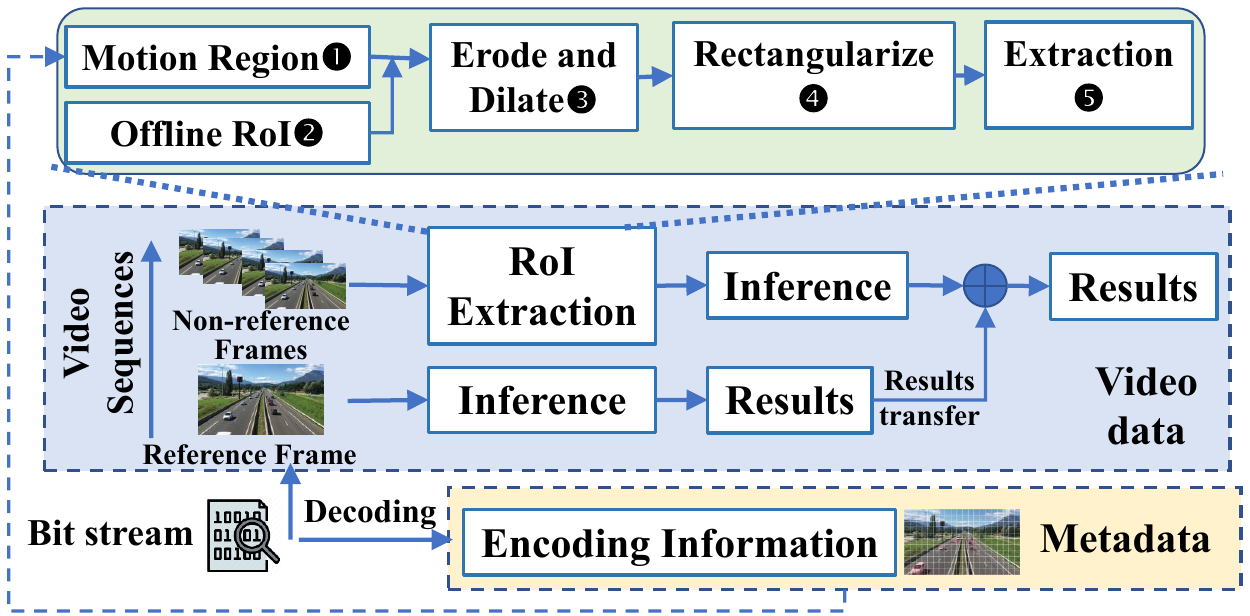}
	\caption{\small Illustration of encoding metadata assisted RoI extraction.}
	\label{fig:roi_extraction}
	\vspace{-0.2in}
\end{figure}
\section{System Design}
\subsection{System Overview}
As depicted in Fig. \ref{fig:system_model}, our video analytics framework adopts an end-edge collaboration paradigm. Video frames are  captured by end camera, then encoded and transmitted to edge node where multiple scales of models are deployed for
analytics. To enhance computational efficiency, we propose a RoI extraction module that leverages encoding metadata to accurately identify RoIs. Considering the constrained processing capacity of edge node, our proposed scheme adaptively allocates computing resources to each model and schedules processing for RoIs, achieving balance between accuracy and processing time. The subsequent sections provide detailed descriptions of the main components of our proposed scheme.
\subsection{ RoI Extraction via Encoding Metadata}
As for video encoding, the input video consists of three frame types: I-frames, P-frames, and B-frames. I-frames are independently encoded and serve as reference frames, while P-frames and B-frames are non-reference frames, encoded using information from the referenced I-frames \cite{yuanyi}. Based on the distinct characteristics of these frame types, we adopt different processing strategies for reference and non-reference frames.
 
 As illustrated in Fig. \ref{fig:roi_extraction}, reference frames are directly analyzed with the most accurate model to ensure high-quality inference. For non-reference frames, we leverage the MVs information from encoded data to extract RoIs through a five-step process. The details are as follows.

\begin{itemize}
	\item Motion region Extraction: We extract MVs from the encoder, retaining macroblocks with nonzero MVs.
	\item Offline non-background regions: Since MVs information is derived through block-matching on pixels, background regions with similar appearances may introduce noise. To address this, we filter out background noise using non-background regions generated offline. Specifically, we apply the widely used semantic segmentation model to generate the regions.
	\item Erode and dilate: The morphological opening operation, involving erosion and dilation, is widely used for noise removal from images. We adopt it to filter isolated points and expand the boundaries of motion regions. The kernel radius is adjustable, allowing us to fine-tune RoI extraction performance.
	\item  Rectangularize: To facilitate image processing, RoIs are adjusted to rectangular shapes. This can be achieved by determining the minimum bounding rectangle that encompasses the RoIs.
	\item  RoI extraction: After the above steps, the generated mask is applied to the original frame for RoI extraction.
\end{itemize}
 
 Notably, this approach is particularly effective in identifying moving objects within motion regions. For stationary object regions, the results can be supplemented by transferring detection results from reference frames, as the positions of static objects remain unchanged.
\subsection{Problem Formulation}
\subsubsection{Video Frame and DNN Model} Denote $\mathcal{F}=$ $\{1, \ldots, f, \ldots, F\}$ by the set of frames to be processed. These frames are divided into $T$ chunks for processing, each of size $c$, where $T=\left\lceil\frac{F}{c}\right\rceil$. For each frame $f$, there are $n_f$ extracted RoIs based on the encoding metadata. The $i$-th RoI of the $f$-th frame is denoted by $b_f^i\left(0 \leq i \leq n_f\right)$.

As for the DNNs, there are $M$ DNNs of different scales available for the analytics task, denote by $\mathcal{M}=$ $\{1,..., m,..., M\}$. Each of them offers various analytics performance and incurs different computational overheads. The extracted RoIs are necessary to be processed by one of these $M$ models. Let $x_{f, m}^i$ be a binary variable indicating whether the RoI $b_f^i$ is analyzed by model $m$. Specifically, $x_{f, m}^i=1$ represents that the RoI $b_f^i$ is analyzed by model $m$. The variable should satisfy
\begin{equation}
\sum_{m=1}^M x_{f,m}^i =1,~~x_{f,m}^i \in \{0,1\},
\label{x1}
\end{equation}
which means that there is one and only one model can be selected for processing for each RoI.

\subsubsection{Accuracy and Latency Model} Denote $a_{f, m}^i$ by the accuracy by $m$-th model for the RoI $b_f^i$. The accuracy $u_f$ of $f$-th frame can be calculated as
\begin{equation}
u_f=\frac{1}{n_f} \sum_{i=1}^{n_f} \sum_{m=1}^M a_{f, m}^i x_{f, m}^i,
\end{equation}

Multiple models of various scales share computing resources of same edge node. Denote $r_{m, t}$ by the allocated computing resources of $m$-th model in $t$-th chunk. The resource allocation for all $M$ models can be denoted by $\mathbf{r}_t=\left\{r_{1, t}, \ldots, r_{m, t}\right\}$. According to \cite{bao2024couple}, the processing latency can be expressed as the function $l\left(r_{m, t}\right)=\frac{\xi_{2, m}}{r_{m, t}+\xi_{1, m}}+\xi_{3, m}$, where $\xi_{1, m}, \xi_{2, m}$ and $\xi_{3, m}$ are constants related to hardware parameters and can be fitted offline. The resource allocation should satisfy overall resource constraint $R_t$, thus we have $\sum_{m=1}^M r_{m, t}=R_t$. The latency of $t$-th chunk can be divided into encoding time $l_1$, transmission time, processing time and queue time, thus we have
\begin{equation}
l_t=l_1+\frac{d_t}{B_t}+\sum_{f=c t}^{c(t+1)} \sum_{i=1}^{n_f} \sum_{m=1}^M\left(x_{f, m}^i \cdot l\left(r_{m, t}\right)+q\left(x_{f, m}^i\right)\right),
\end{equation}

The objective is to optimize the variable RoI scheduling $\mathbf{x}_{\mathbf{f}, \mathbf{m}}$ and computing resource allocation $\mathbf{r}_t$, to maximize the analytic accuracy and reduce the processing latency.
\begin{align}
\mathcal{P}_0:& \max_{\mathbf{x_{f,m}}, \mathbf{r}_t} \quad \omega_1 \cdot  \frac{1}{F}\sum_{f} u_f - \omega_2 \cdot \frac{1}{T}\sum_{t=1}^{T} (l_t - L_t), \label{eq:1}
\\\text{s.t.} &~~~ ~~~ \eqref{x1}, ~~~ \sum_{m=1}^M r_{m,t} = R_t, \forall t \in \{1,...,T\}\nonumber,
\end{align}
The weights $\omega_1$ and $\omega_2$ control the trade-off between accuracy and processing latency. The problem is a mixed integer nonlinear programming problem (MINLP), which is NP-hard. Meanwhile, it involves both discrete and continuous variables, which increases the difficulty of solving the problem.

\subsection{Algorithm Design}
\subsubsection{Computing Resource Allocation} Based on the observation in \cite{wang2024joint}, the problem can be solved by binary variable relaxation. We relax the discrete variable $\mathbf{x}_{\mathbf{f}, \mathrm{m}}$ to continuous variable $\widetilde{\mathbf{x}_{\mathbf{f}, \mathbf{m}}}=\left\{\widetilde{x_{f, m}^i}, i \in\left[1, n_f\right]\right\}$. The relaxed form of $l_t$ can be formed as
\begin{equation}
\tilde{l_t} = l_1 + \frac{d_t}{B_t} + \sum_{f=ct}^{c(t+1)} \sum_{i=1}^{n_f} \sum_{m=1}^{M} (\widetilde{x_{f,m}^i} \cdot l(r_m) +q(\widetilde{x_{f,m}^i} )),
\label{eq:l_t}
\end{equation}
The problem $\mathcal{P}_0$ can be relaxed to
\begin{align}
\mathcal{P}_1:& \max_{\mathbf{\widetilde{x_{f,m}}}, \mathbf{r_t}} \quad \omega_1 \cdot  \frac{1}{F}\sum_{f} u_f - \omega_2 \cdot \frac{1}{T}\sum_{t=1}^{T} (\tilde{l_t} - L_t), \label{eq:1}
\\\text{s.t.} &~~~ 0 \leq \widetilde{x_{f,m}^i} \leq 1 , ~~~ \eqref{x1}, ~~~ \sum_{m=1}^M r_{m,t} = R_t, \forall t \in \{1,...,T\}\nonumber,
\end{align}

Notably the problem related to variable $r_{m, t}$ is convex. The value of $r_{m, t}$ can be obtained by KKT condition. According to the convex theorem, we first derive the Hessian matrix of the expression $\tilde{l}_t$ with respect to variable $r_{m, t}$.
\begin{equation}
\begin{aligned}
H  =\frac{\partial^2 \widetilde{l}}{\partial r_{m} \partial r_{m ^\prime}} 
= \begin{cases}\frac{2 \xi_{2,m}}{(r_{m}+ \xi_{1,m})^3 } & \text { if } m = m ^\prime\\
0  &\text { otherwise. }\end{cases}
\end{aligned}
\end{equation}
\begin{figure}[t]
	\centering
	\subfigure[]{
		\includegraphics[width=0.23\textwidth]{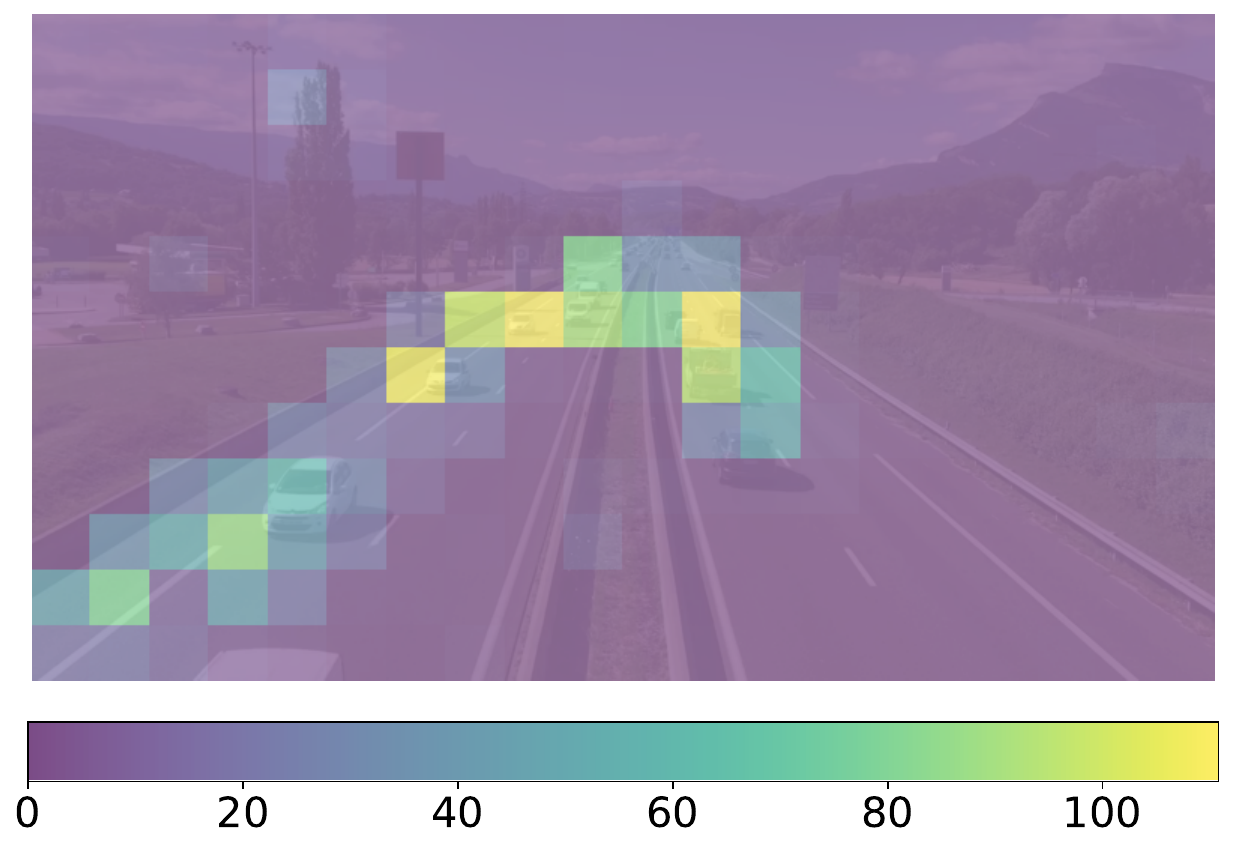}\label{fig:bitrate_allocation}}
	\subfigure[]{
		\includegraphics[width=0.23\textwidth]{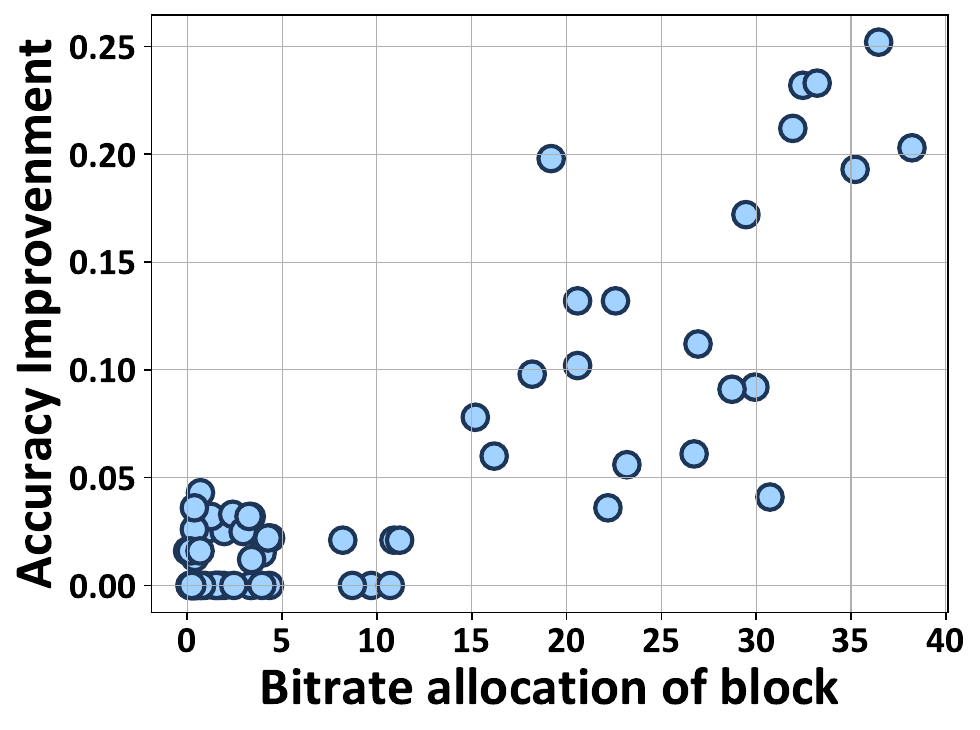}\label{fig:bitrate_result}}
		\vspace{-0.1in}
	\caption{The relationship between the accuracy improvement and bitrate allocation of block: The heatmap of (a) accuracy improvement of each block, (b)average bitrate allocation of each CTU.}
	\label{fig:block_bitrate}
	\vspace{-0.2in}
\end{figure}
Since $\xi_{2, m}$ is positive, the value of $H$ is non-negative.

\subsubsection{RoI Scheduling Algorithm Design} 
\begin{figure}[t]
	\label{alg:ORS}
	\renewcommand{\algorithmicrequire}{\textbf{Input:}}
	\renewcommand{\algorithmicensure}{\textbf{Output:}}
	\begin{algorithm}[H]
		\caption{RoI Scheduling Algorithm Design}
		\label{alg:ORS1}
		\begin{algorithmic}[1]
			\REQUIRE The queue state $q(t)$, processing rate $\mu$, model set $\mathcal{M}$
			\ENSURE  The scheduling decisions {$x_{f,m}^i$}
			
			\STATE According to the allocated bitrate in each RoI, and divide RoIs into $M$ group, denoted by $G=\{G_1,...,G_M\}$
			\WHILE {$\mathcal{M} \neq \varnothing$}
			\FOR {$G_r \in G$}
			\IF{$G_r$ is empty} 
			\STATE $\mathcal{M} \leftarrow  \mathcal{M}\backslash{r}$ 
			\ENDIF
			\ENDFOR
			\STATE Minimum workload queue set $\Gamma \leftarrow \arg\min_{m \in \mathcal{M}} \frac{q_m}{\mu_m}$
			\\// {Case: The workloads of all queues are equal.}
			\IF{$\left|\Gamma \right| ==\left| \mathcal{M} \right| $}
			\STATE $\beta =  \min_{\{r \in  \mathcal{M}\}}{\left| G_r \right|}$    //groups with least workload 
			\STATE $q_r = q_r +\beta$ and Update $G_r$, $\forall r \in \Gamma$
			\STATE break
			\ENDIF
			\FOR{$r \in \Gamma$}
			\STATE $q_r = q_r + \min{(( \min_{\{m \in \mathcal{M}\backslash{\Gamma}\}}{\frac{q_m}{\mu_m}}-\frac{q_r}{\mu_r}) \cdot \mu_r, \lvert G_r \rvert)}$
			\STATE Update $G_r$ 
			\ENDFOR
			\ENDWHILE
			\IF{there exist $G_r$ not equal zero} 
			\STATE Randomly initialize the scheduling decisions for not assigned RoIs. Computing utility function $U$
			\WHILE{$T<T_{max}$}
			\STATE Generate a new decision and recalculate utility $U'$
			\STATE $\eta \leftarrow \frac{1}{1+e^{\left(\frac{\hat{U'}-U}{\tau}\right)}}$, $U \leftarrow U'$
			\ENDWHILE
			\ENDIF
		\end{algorithmic}
	\end{algorithm}
	\vspace{-0.3in}
\end{figure}
To facilitate RoI scheduling, it is necessary to estimate the accuracy improvement of RoIs when processed with a larger model. Direct calculation of this value requires processing the RoI twice. Such a chicken-egg paradox makes us search for an indicator. We contend the bitrate allocated to each Coding Tree Unit (CTU) during encoding can serve as an indicator by experiment. Each scatter point in Fig. \ref{fig:bitrate_result} represents the accuracy improvement value of RoIs and average bitrate of the covered CTUs. When the bitrate surpasses a certain threshold (about $12$kbps in Fig. \ref{fig:bitrate_result}), the accuracy improvement value is almost positively correlated with the bitrate. This is because the allocated bitrate reflects the complexity and variability of video content. According to the bitrate allocation information, the RoIs are divided into $M$ group, denoted by $G=\left\{G_1, \ldots, G_M\right\}$.

The RoI scheduling algorithm is shown in Alg. \ref{alg:ORS1}. Based on the RoI grouping results, the design intuition is to prioritize scheduling RoIs to models of the appropriate complexity, while ensuring the workload of multiple models as balanced as possible. Considering the varying processing speeds $\mu_m$ of models, the workload of model $m$ is defined by the expected completion time $\frac{q_m(t)}{\mu_m}$. In each iteration, we retain the corresponding models that still have RoIs to be processed in that group, as shown in lines 3-7. And then we select the set of models with the lowest workload in line 8. The workload of these models will be increased to match the next lowest workload of the other models, as shown in lines 9-17. The iteration will terminate when the workload across all models is equal or when all RoIs are analyzed. If there still exist unallocated RoIs, an Markov-based approach \cite{zhang2021adaptive} is adopted to find the scheduling decision. The utility function is defined as objective function value. The optimal value will be searched for the appropriate value after $T_{\max }$ searches.

Finally, we analyze complexity of the algorithm. The time complexity of first stage is $O(M \log M)$, dominated by workload sorting of models. As for the second stage, assuming that $L$ iterations are required for convergence, the time complexity is $O(n L M)$, where $n$ represents the number of RoIs assigned in this stage. The number of models $M$ and the number of RoIs $n$ are relatively small. The value of $L$ will not exceed $T_{\max}$. Thus the overall algorithm complexity remains acceptable.

\section{Performance Evaluation}
\subsection{Evaluation Setup}
\textbf{Platforms.} In the experiment, we use two computing platforms to deploy our scheme. One is NVIDIA Jetson Tx2 with 256 CUDA cores. The other is edge server including NVIDIA GeForce RTX 3080Ti GPU and Intel(R) Core(TM) i9-11900K@ 3.50GHz CPU. The video encoder is based on FFmpeg tool for H.265 encoding. The encoding metadata is extracted by an opensource MVs extractor tool implemented by C++ language\footnote{Motion Vector Extractor, https://github.com/LukasBommes/mv-extractor. Accessed March 24, 2025.}.

\textbf{Task and metrics.} We evaluate our proposed scheme on object detection task. For detection model, we select three models with different computing capacity, including Yolov8s, Yolov8m and Yolov8x model. F1 score, which is defined as the harmonic mean of precision and recall, is used to measure the analytic accuracy.

\textbf{Datasets.} To evaluate the effectiveness of our proposed scheme, we select two real-world static video datasets (`Relaxing highway traffic' and `Road traffic video') from Yoda dataset. The resolutions of the two videos are $1280 * 720$ and the frame rates are $30$ fps.


\textbf{Baselines.} We evaluate the scheme with these baselines.
\begin{itemize}
\item \textbf{Inference the entire frame:} All video frames will be inferenced by the Yolov8x model.
\item \textbf{AccDecoder-based} \cite{yuan2023accdecoder}: The denoised coded information MVs are used for detection results reuse.
\item \textbf{ELF-based} \cite{zhang2021elf}: The RoI extraction approach is based on the attention-based LSTM neural network and previous results. Then the extracted RoI are scheduled to multiple models for processing.
\end{itemize}
\subsection{ Comparison of Analytic Accuracy}

\begin{table}[t]
	\begin{center}
		\caption{Latency performance of RoI extraction scheme comparison} \label{tab:cap}
		\begin{tabular}{c|c|c}
			\hline\hline
			Scheme & Nvidia GeForce RTX 3080Ti & Nvidia Jetson Tx2
			\\
			\hline
			Ours     & 1.26ms     & 4.52ms       \\
			\hline
			ELF-based    & 6.16ms     & 14.67ms       \\
			\hline
			CAM-based & 3.28ms & 6.96ms\\
			\hline\hline
		\end{tabular}
	\end{center}
	\vspace{-0.1in}
\end{table}

\begin{figure}[t]
	\subfigure[]{
		\includegraphics[width=0.23\textwidth]{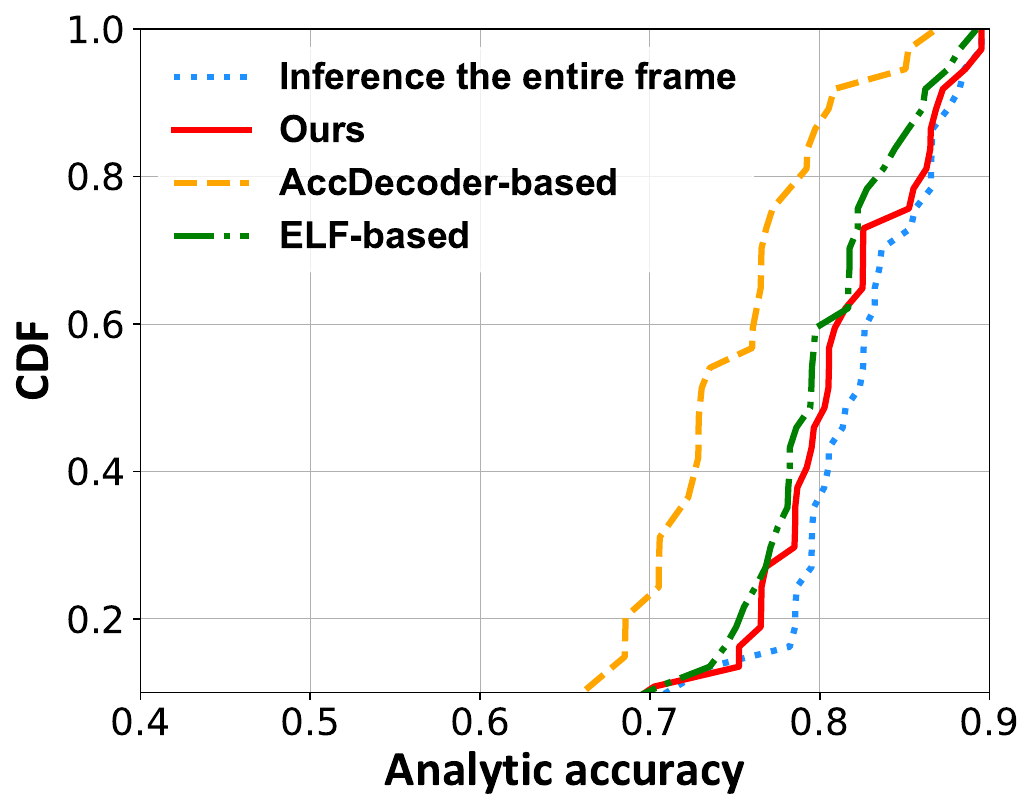}\label{fig:1}}
	\subfigure[]{
		\includegraphics[width=0.23\textwidth]{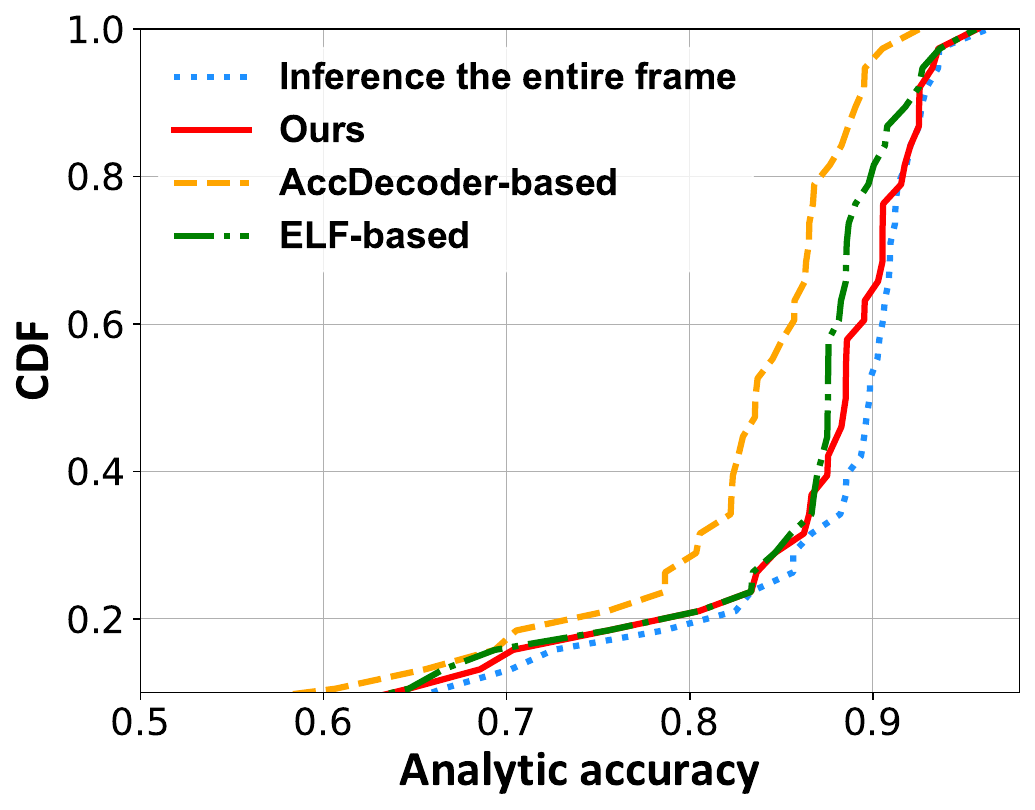}\label{fig:2}}
	\vspace{-0.1in}
	\caption{Comparison of accuracy among algorithms on dataset (a) 1, (b) 2.}
	\label{fig:motivation}
	\vspace{-0.2in}
\end{figure}

Figure \ref{fig:1} and \ref{fig:2} shows the accuracy performance of four schemes on two datasets. The results illustrate that the accuracy of our scheme only reduce by $1.6 \%$, compared to the Inference the entire frames scheme. This slight reduction confirms the effectiveness of our design for RoI extraction and adaptive model inference. The accuracy of our scheme is slightly better than that of ELF scheme, because scheduling scheme in ELF does not take into account the detection difficulty of RoIs, but only the impact of workload. The accuracy of AccDecoder scheme is relatively poor, due to the direct use of less accurate encoded information to reuse detection results. Overall, our proposed scheme can improve accuracy by $2.2 \%$ on average, compared to the two benchmarks.

\begin{figure}[t]
	\centering  
	\subfigure[]{
		\includegraphics[width=0.15\textwidth]{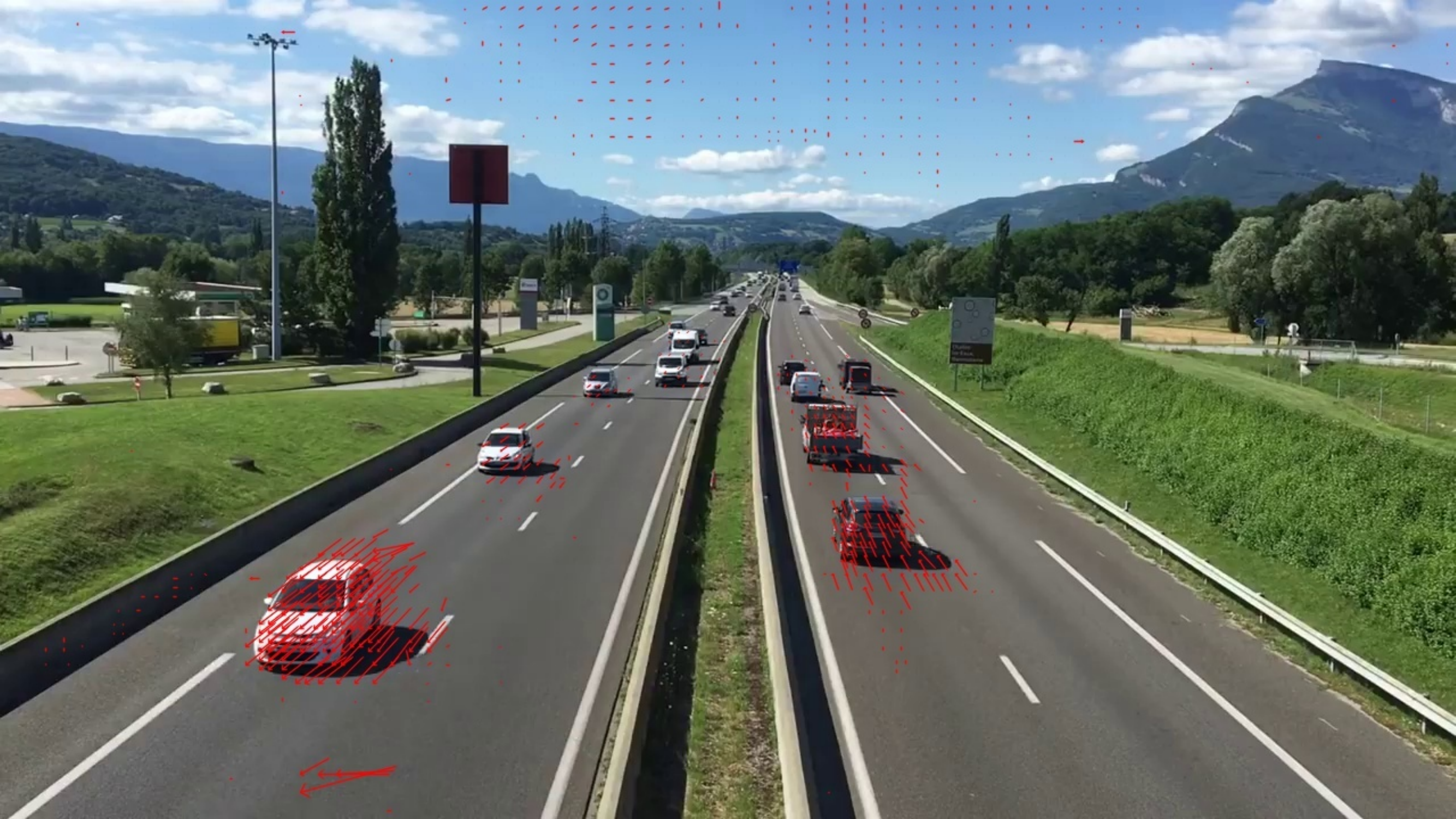}\label{fig:original_frame}}
	\subfigure[]{
		\includegraphics[width=0.15\textwidth]{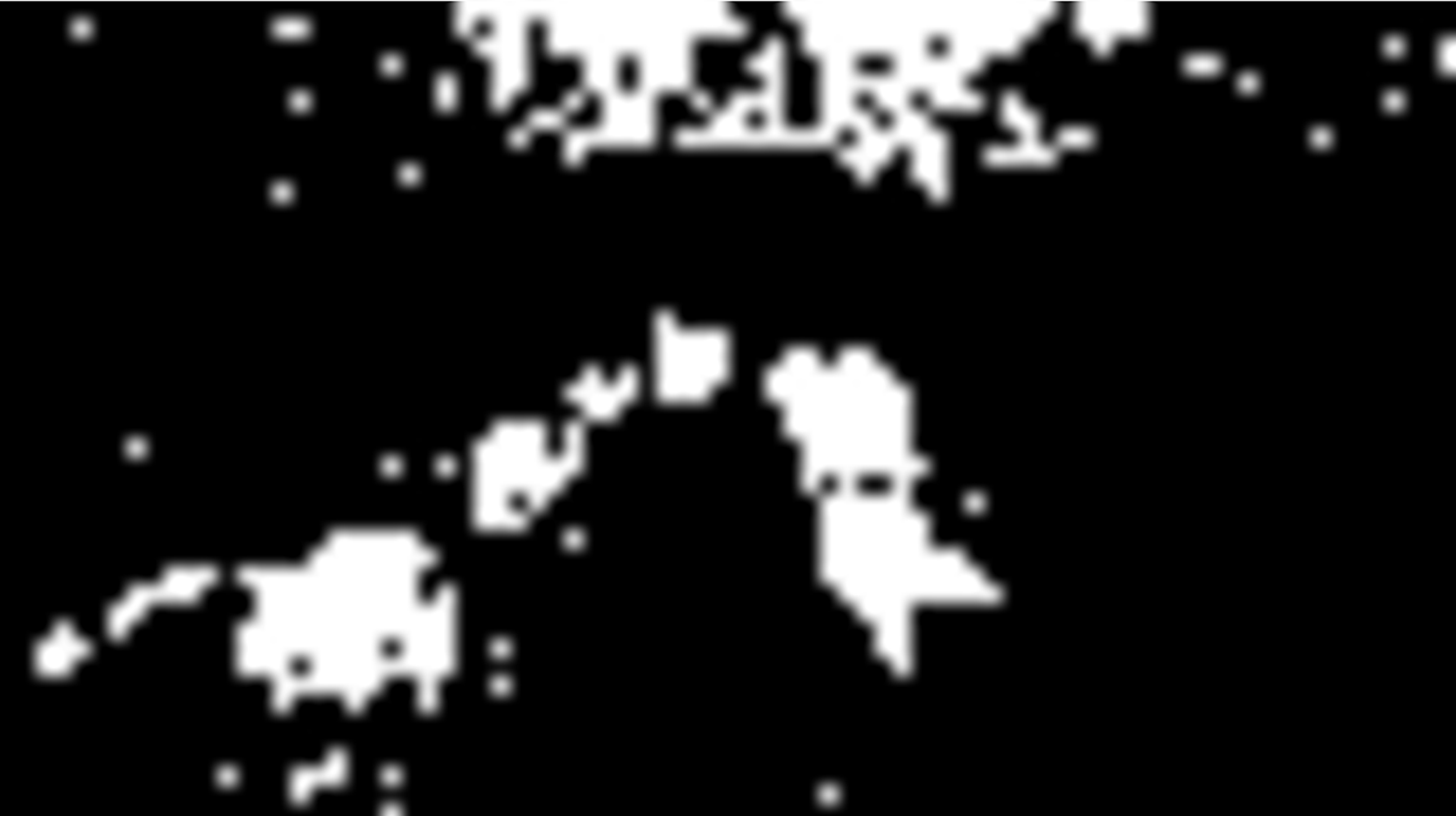}\label{fig:binary_image}}
	\subfigure[]{
		\includegraphics[width=0.15\textwidth]{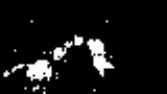}\label{fig:binary_image_after}}
	\subfigure[]{
		\includegraphics[width=0.15\textwidth]{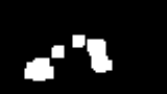}\label{fig:isolated}}
	\subfigure[]{
		\includegraphics[width=0.15\textwidth]{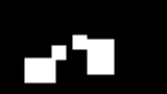}\label{fig:rect_image}}
	\subfigure[]{
		\includegraphics[width=0.15\textwidth]{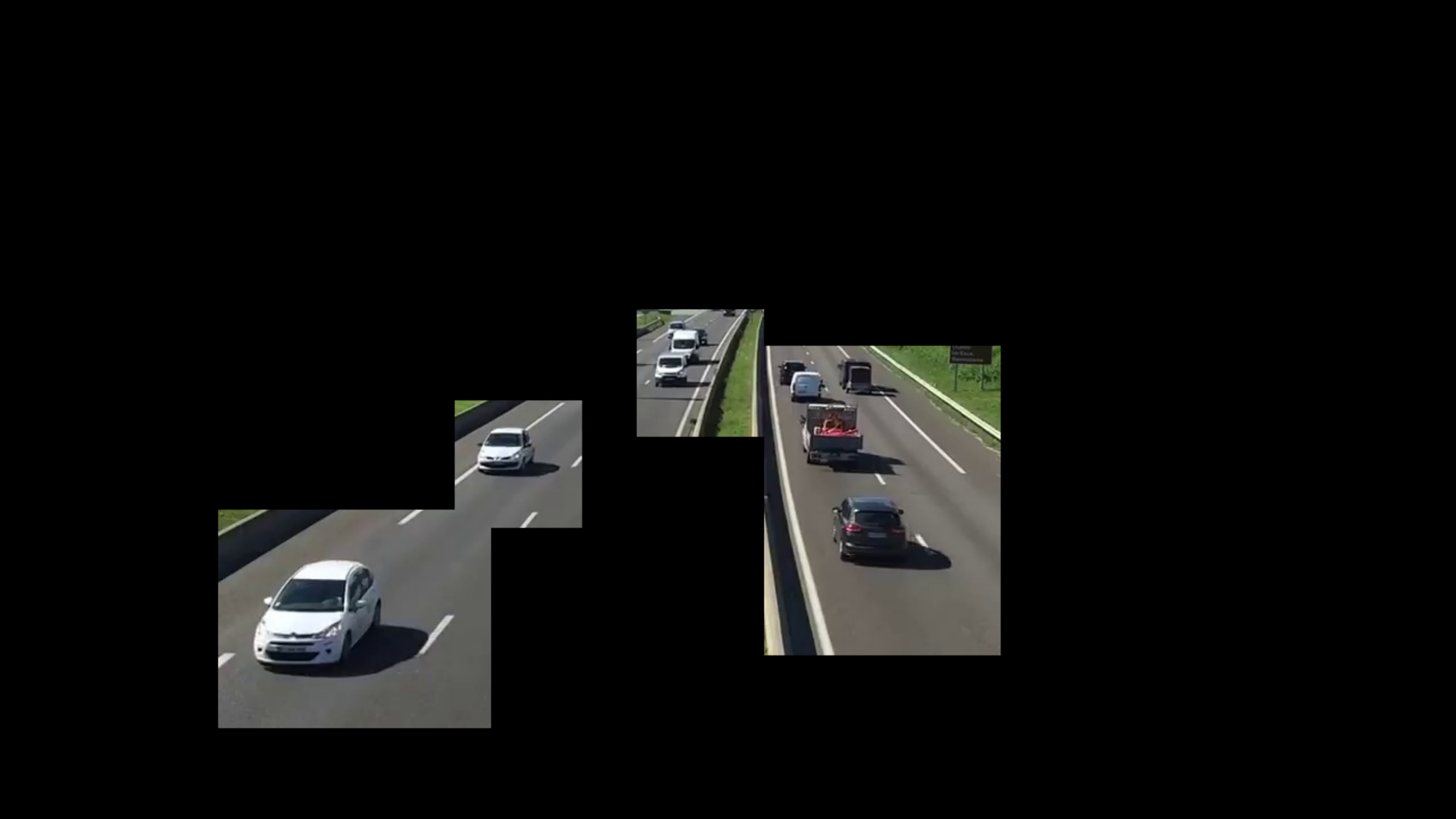}\label{fig:img}}
	\caption{An example of the RoI extraction visualization of five steps: (a) frame with MVs, (b)-(e) represent the results of steps 1-5 respectively.}
	\label{fig:visualization}
\end{figure}

\begin{figure}[t]
	\begin{minipage}[t]{0.23\textwidth}
		\centering
		\includegraphics[width=\textwidth]{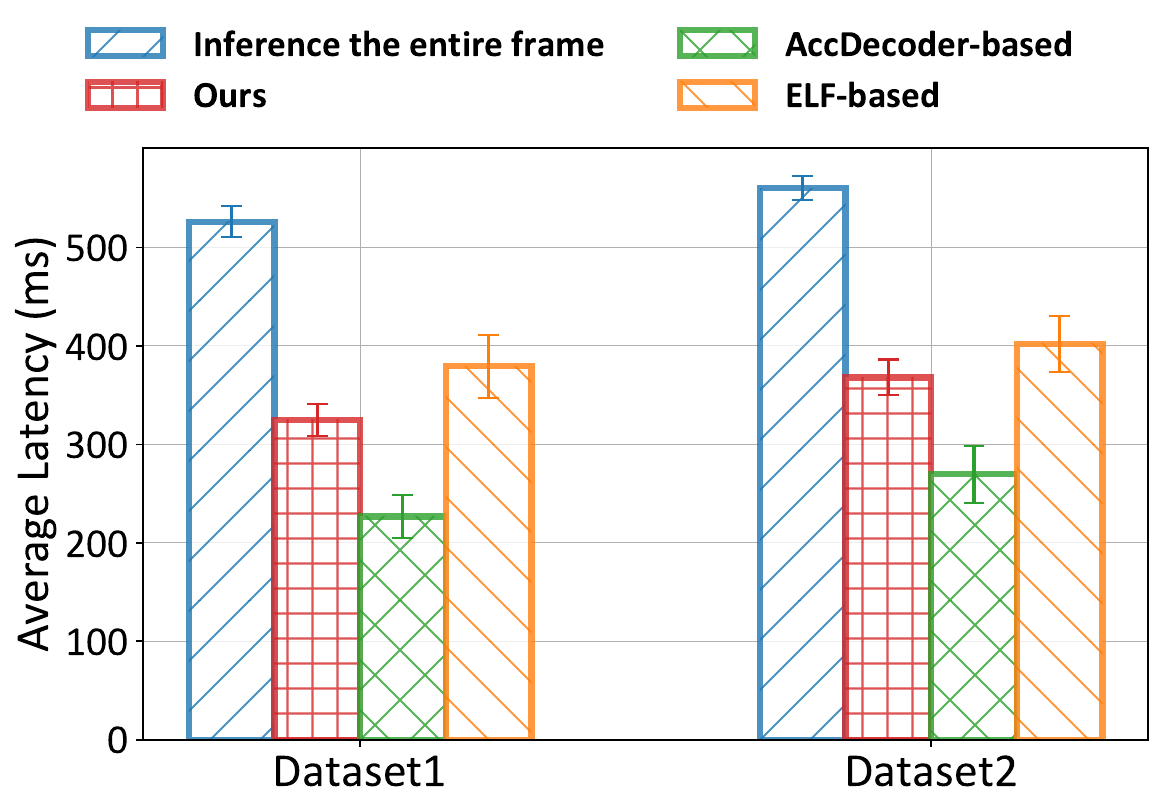}
		\caption{Comparison of latency among algorithms.}
		\label{fig:latency}
	\end{minipage}
	\hspace{0.0051in}
	\begin{minipage}[t]{0.26\textwidth}
		\centering
		\includegraphics[width=\textwidth]{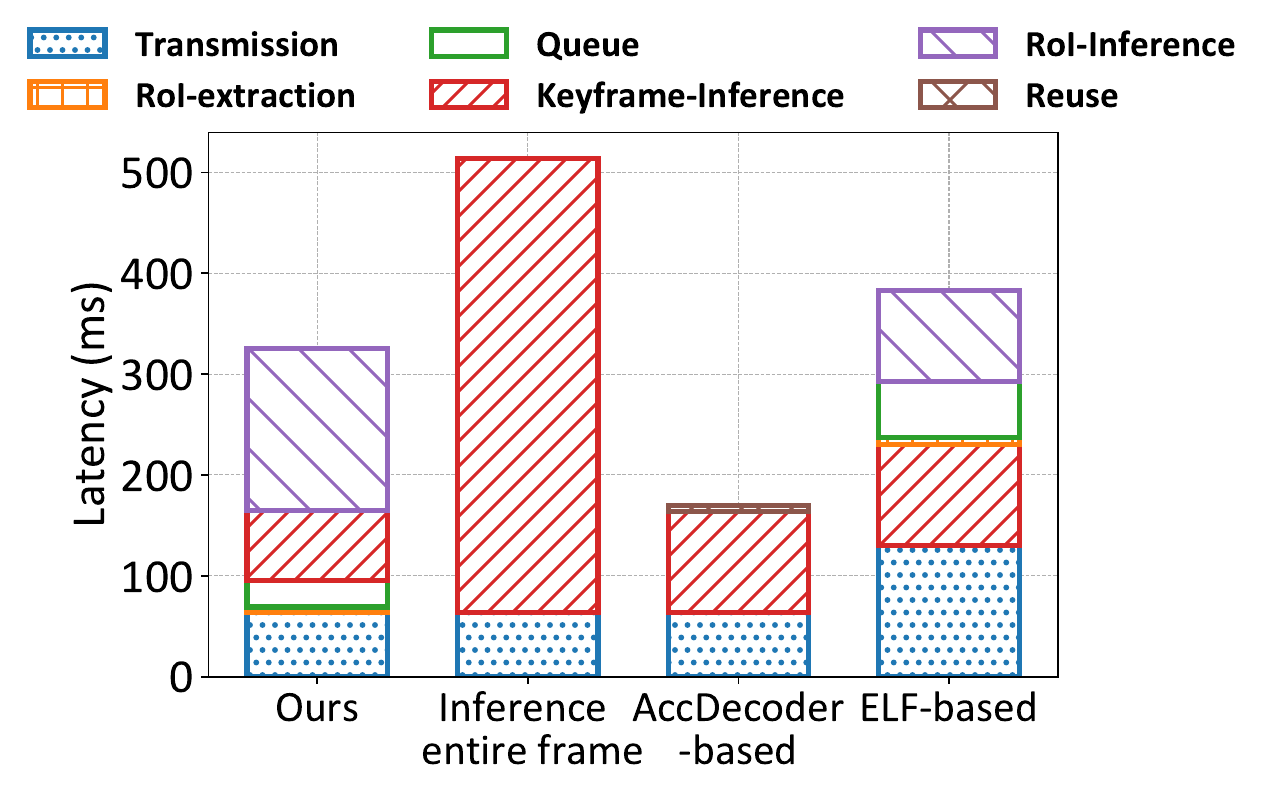}
		\caption{Decomposition of latency among algorithms.}
		\label{fig:Decomposition of latency}
	\end{minipage}
	
\end{figure}

\subsection{Comparison of Latency}
\subsubsection{Latency Comparison of RoI Extraction Scheme} Fig. \ref{fig:visualization} shows the visualization of the intermediate and final results on one typical frame of five steps described in Section III-B. At the same time, we compare the latency performance with two latest RoI extraction schemes \cite{cheng2023edge, zhang2021elf} across two hardware environments, with the experimental results shown in Table \ref{tab:cap}. The results indicate that RoI extraction scheme proposed in our paper incurs minimal overhead on both hardware platforms.
\subsubsection{Average Latency Comparison} We evaluate the average latency of four schemes on two datasets. The results are shown in Fig. \ref{fig:latency}. As depicted, compared to Inference the entire frame scheme, our scheme has a significant latency reduction by approximately $40 \%$ on both two datasets. The reduction comes from the fact that RoI extraction is performed with less overhead, reducing inference overheads. Compared to the ELF-based scheme, the latency reduction of our proposed scheme is up to $18 \%$. It is worth noting that the AccDecoder-based scheme performs optimally in terms of latency. The reason results from the detection overhead saved by reusing the detection results of the previous frame. However, this reuse will result in a reduction in the detection accuracy, which can be concluded from Fig. \ref{fig:latency}. The motion vectors in coded information are only pixel-point matches, and their direct use for the semantic information produces errors.
\subsubsection{Latency Breakdown} The latency breakdown in one typical chunk is shown in Fig. \ref{fig:Decomposition of latency}. The transmission delay of ELF framework is larger than others, because the other schemes transmit encoded bit streams, whereas ELF transmits the extracted RoIs. The task inference time basically accounts for more than half of overall latency, which represents that extracting RoIs to reduce inference latency is essential. Compared to Inference the entire frames scheme, our scheme has significantly lower reference frame inference latency.
\section{Conclusion}

In this paper, we have presented a cost-effective scheme for extracting RoIs and adaptive inference by leveraging informative encoding metadata. Based on MVs information, we have proposed an RoI extraction scheme to accurately detect RoIs with movement within static background. To analyze these RoIs efficiently and accurately, we have proposed an algorithm for computing resource allocation and RoIs scheduling. Extensive experimental results have shown that our scheme reduces latency by nearly $40 \%$ and improves $2.2 \%$ on average in accuracy, outperforming other benchmarks.

\bibliographystyle{IEEEbib}
\bibliography{ref}

\end{document}